\documentclass[amsmath, amssymb, aps, prb, twocolumn, notitlepage, superscriptaddress]{revtex4-1}

\usepackage[utf8]{inputenc}
\usepackage{graphicx}               
\usepackage[caption=false]{subfig}  
\usepackage{dcolumn}                
\usepackage{bm}                     
\usepackage[dvipsnames]{xcolor}     
\usepackage{slashed}                
\usepackage{hyperref}               
\usepackage{enumitem}               
\usepackage{mathtools}              

\newcounter{myparagraphs}

\begin{document}
\title{Beyond Ohm's law - Bernoulli effect and streaming in electron hydrodynamics}
\author{Aaron Hui}
\affiliation{School of Applied \& Engineering Physics, Cornell University, Ithaca, New York 14853, USA}

\author{Vadim Oganesyan}
\affiliation{Department of Physics and Astronomy, College of Staten Island, CUNY, Staten Island, NY 10314, USA}
\affiliation{Physics Program and Initiative for the Theoretical Sciences,
The Graduate Center, CUNY, New York, NY 10016, USA}

\author{Eun-Ah Kim}
\affiliation{Department of Physics, Cornell University, Ithaca, New York 14853, USA}

\date{\today}

\begin{abstract}
Recent observations of non-local transport in ultraclean 2D materials raised the tantalizing possibility of accessing hydrodynamic correlated transport of many-electron state.
However, it has been pointed out that non-local transport can also arise from impurity scattering rather than interaction. 
At the crux of the ambiguity is the focus on linear effects, i.e. Ohm's law, which cannot easily differentiate among different modes of transport.
Here we propose experiments that can reveal rich hydrodynamic features in the system by tapping into the non-linearity of the Navier-Stokes equation. 
Three experiments we propose will each manifest unique phenomenon well-known in classical fluids: the Bernoulli effect, Eckart streaming, and Rayleigh streaming. 
Analysis of known parameters confirms that the proposed experiments are feasible and the hydrodynamic signatures are within reach of graphene-based devices.  
Experimental realization of any one of the three phenomena will provide a stepping stone to formulating and exploring the notions of nonlinear electron fluid dynamics with an eye to celebrated examples from classical non-laminar flows, e.g. pattern formation and turbulence.

\end{abstract}

\maketitle
\section{Introduction}

Electron hydrodynamics offers a powerful framework to understand transport in strongly correlated electron systems.
\cite{Damle1997, Son2007, Sachdev2009, Hartnoll2007, Fritz2008, Foster2009, Muller2009, Davison2014,Lucas2015,Lucas2017,Zaanen2019, Torre2015a, Levitov2016, Varnavides2020}
The pursuit of electron hydrodynamics gained new impetus with the advent of recent experiments in a number of ultraclean 2D materials\cite{Crossno2016, Bandurin2016, Kumar2017, Ku2019, Sulpizio2019, Moll2016, Gooth2018, Gusev2018, Levin2018, Gusev2020} making a case for electron hydrodynamics through observations of non-local transport, consistent with viscous flows familiar in classical fluids. The observations such as vortices, Poiseuille-like flow profiles, and unconventional channel width dependencies of resistance are indeed consistent with viscous effects in a linearized Navier-Stokes equation.
However, these results are all in the linear response regime and can be ultimately described using a non-local variant of Ohm's law.
Indeed, the linearized Navier-Stokes equation can be simply recast using a non-local conductivity $\sigma(q)$\cite{Pellegrino2016,Shytov2018,Hui2020}. While non-local transport can certainly be couched in the formalism of hydrodynamics, it is also clear that inherently  finite length scales of a realistic fermionic system can conspire to produce non-local transport indistinguishable from that implied by the Navier-Stokes equation\cite{Hui2020}.
Other ways of accessing electron hydrodynamics are of great interest as we seek to understand and isolate competing effects.

\begin{figure*}
\label{fig:1}
    \centering
    \begin{subfloat}[][]{
        \includegraphics[scale=.5]{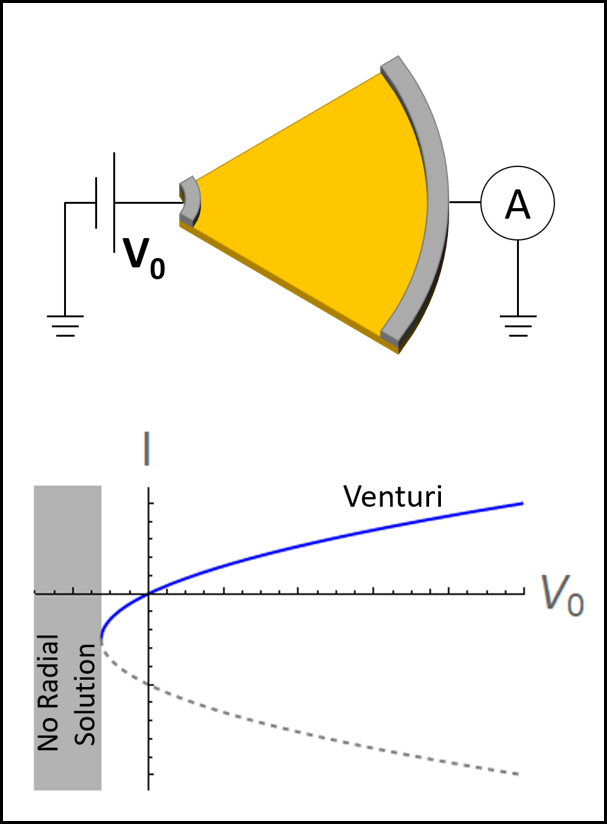}
        \label{fig:venturi experiment}
    }
    \end{subfloat}
    \quad
    \begin{subfloat}[][]{
        \includegraphics[scale=.5]{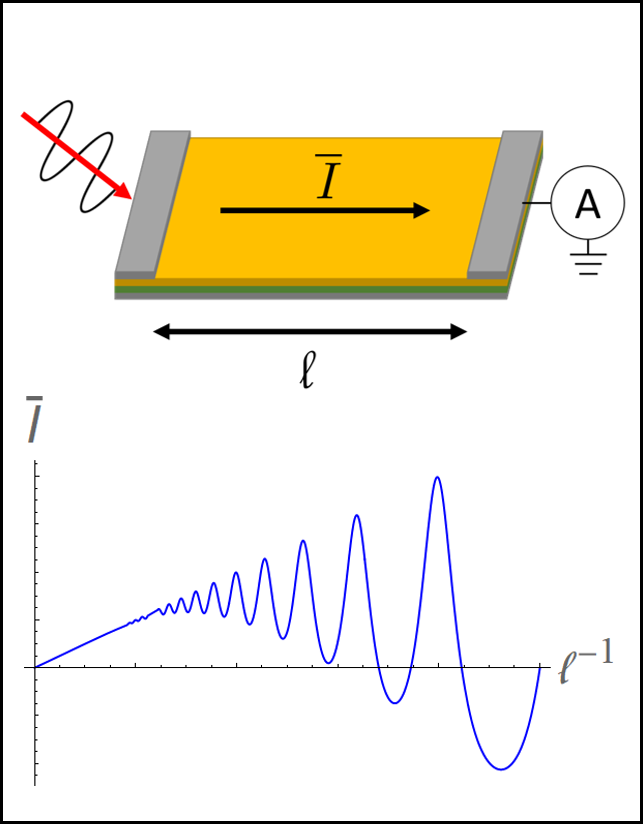}
        \label{fig:eckart experiment}
    }
    \end{subfloat}
    \quad
    \begin{subfloat}[][]{
        \includegraphics[scale=.5]{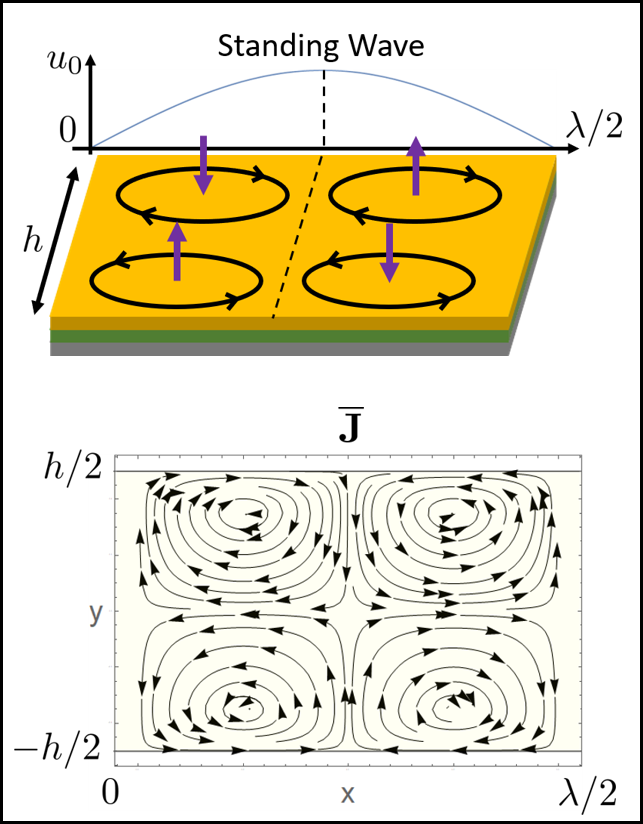}
        \label{fig:rayleigh experiment}
    }
    \end{subfloat}
    \label{fig:experimental proposals}
    \caption{Proposed experimental setups and sketches of their observed effects. 
    a) The Venturi geometry, comprised of a circular wedge of the hydrodynamic material in yellow. A nonlinear I-V characteristic with $I \sim \sqrt{V_0}$ behavior is expected, marked in blue. The gray dashed line represents an unstable solution branch, while the gray region represents a possible instability towards turbulent and/or intermittent flow.
    b) Eckart streaming. A voltage oscillation of zero mean is driven on one side of a back-gated device, leading to a rectified DC current $\overline{I}$.
    For large $l$, the DC current scales as $l^{-1}$. 
    For small $l$, oscillations due to interference with the reflected wave become visible.
    c) Rayleigh streaming. In a similar back-gated geometry of (b), a standing wave of current oscillations of amplitude $u_0$ and of period $\lambda$ along $x$ is imposed, leading to an oscillating magnetic field pattern of period $\lambda/2$ along x. These magnetic fields arise due to the formation of vortical current cells of size $\lambda/4$ along x and $h/2$ along y, shown in the lower panel.}
\end{figure*}

The overarching goal of this paper is to highlight the existence of nonlinear electron \emph{phenomena} that may be associated with an effective hydrodynamic description. 
With that in mind, we directly adapt the Navier-Stokes (NS) equations of classical fluid dynamics to make the discussion of the electron phenomenology explicit. 
We do not tackle the important and difficult question of a proper microscopic derivation of NS -- indeed, there is evidence that many available electron devices are not quite in the asymptotic hydrodynamic regime\cite{bandurin2018fluidity, Principi2019}. 
We do, however, find strong evidence in known material and device parameters to support feasibility of our proposals.  It is worth emphasizing that while the phenomena we focus on in this work are leading deviations from linear response, the NS results we obtain also suggest the presence of instabilities at finite non-linearity. 
As in traditional classical hydrodynamics, these different regimes are naturally demarcated using dimensionless Reynolds numbers.

In Fig.~\ref{fig:1}, we summarize the three proposals that we discuss in this paper.
The rest of the paper is organized as follows. 
Section \ref{sec:2setup} sets up the notation and formalism of NS, paying particular attention to the spectrum of Reynolds numbers required to quantify nonlinear phenomena. 
Here, we also collect Reynolds number estimates from known parameters for graphene. Section III focuses on the manifestation of the Bernoulli effect in the nonlinear current-voltage response of an electron funnel. Section IV derives the generation of downconverted DC current from a localized finite-frequency excitation, analogous to Eckart streaming or "quartz wind". 
Section V describes the generation of static electron vortices (akin to Rayleigh streaming) from an extended AC excitation. Sections II-V are accompanied by Appendices A-D containing complete details of calculations.
Finally, we close with a summary of results and a discussion of open problems, including the role of interactions.

\section{Formalism and Parameters}
\label{sec:2setup}
\subsection{Equations of fluid dynamics}

The hydrodynamics of an electron fluid, as a long-wavelength effective theory, is described by a set of conservation laws for variables which decay slowly compared to the coarse-graining scale of the system.
The momentum (Navier-Stokes) and density continuity equations, which will be our primary interest in this paper, are\footnote{The curl is interpreted in 3D, so that it sends vectors to vectors.}
\begin{align}
    \frac{\partial n}{\partial t} + \nabla\cdot &(n\mathbf{v}) = 0 
    \label{eq:density continuity}
    \\
    \frac{\partial (\rho \mathbf{v})}{\partial t} =& \mathbf{F}_\text{conv} -\nabla p - \rho_e \nabla \phi
    \nonumber \\
    & + \left[\frac{4}{D}\nu + \tilde{\zeta}\right] \rho \nabla \nabla \cdot \mathbf{v} - \rho\nu \nabla \times\nabla\times \mathbf{v} - \rho \gamma \mathbf{v} 
    \label{eq:Navier-Stokes}
    \\
    \mathbf{F_\text{conv}} \equiv& -\nabla\cdot(\rho \mathbf{v}\otimes\mathbf{v}) = \rho\mathbf{v}\cdot\nabla\mathbf{v} + \mathbf{v}\nabla\cdot(\rho\mathbf{v})
    \label{eq:F_conv}
\end{align}
where $\mathbf{v}$ is the velocity field, $n$ is the number density field with particles of mass $m$ and charge $e$ ($\rho$ and $\rho_e$ are the mass and charge densities, respectively). The convective term $\mathbf{F}_\text{conv}$ is written to emphasize that it acts as an effective force; this will be the primary source of nonlinear behavior.
The remaining terms may also be thought of as (generalized) forces, and we can take their ratios for a particular flow pattern to characterize their relative importance.
In addition to the conventional ``viscous" Reynolds number $\operatorname{Re}_\nu$ corresponding to shear dissipation, a momentum-relaxation Reynolds number $\operatorname{Re}_\gamma$ will be of interest.
For simple non-singular flow profiles these may be expressed
\begin{align}
    \operatorname{Re}_\nu \equiv& \frac{\nabla \cdot (\rho\mathbf{v}\otimes \mathbf{v})}{\rho\nu \nabla^2 \mathbf{v}} = \frac{v L}{\nu} = \frac{IL}{\rho_e h\nu}
    \label{eq:Re_nu}
    \\
    \operatorname{Re}_\gamma \equiv& \frac{\nabla \cdot (\rho\mathbf{v}\otimes \mathbf{v})}{\rho\gamma \mathbf{v}} = \frac{v}{L\gamma} = \frac{I}{\rho_e h L\gamma}
    \label{eq:Re_gamma}
\end{align}
with help of characteristic velocity $v$, gradient $1/L$, channel width $h$ and net current $I = \rho_e h v$.
In this paper, we primarily focus on the limit of low Reynolds numbers $\operatorname{Re}_\gamma, \operatorname{Re}_\nu \ll 1$, i.e. leading corrections to linear response\footnote{The third dimensionless number which captures the relative strength of pressure (and potential) terms to convection turns out to be related to the Mach number.}.

Following standard practice, we make a further assumption of local equilibrium to write equations of state for $p$ and $\phi$ which closes the set of continuity equations above.
We take a back-gated geometry as shown in Fig.~\ref{fig:eckart experiment}, where the hydrodynamic metal and the backgate separated by a distance $d$ have a capacitance per unit area $C=\frac{\epsilon \epsilon_0}{d}$.
Therefore, we take the following local relationships
\begin{align}
    p =& s_\text{FL}^2 \rho \label{eq:pressure eq of state}\\
    \phi =& \rho_e/C \label{eq:capacitance eq of state}
\end{align}
where $s_\text{FL}$ is a constant corresponding to the speed of sound in an uncharged, undamped fluid (i.e. a Fermi liquid).
In Eq.~\ref{eq:capacitance eq of state}, also called the ``gradual channel approximation,'' the long-range Coulomb tail is screened by the gate so that the longitudinal dispersion is gapless.
This approximation is valid when the distance $d$ between the hydrodynamic metal and the gate is much smaller than the typical wavelength of oscillations.\cite{Dyakonov1993, Tomadin2013,Torre2015}
Therefore, both $p$ and $\phi$ obey the same functional form; if the density $\rho=\rho^{(0)}$ is constant, $p$ can be absorbed into an effective voltage $\phi_\text{eff} \equiv \phi + \frac{p}{\rho_e^{(0)}}$ in the momentum equation.
In particular, as a result of Eq.~\eqref{eq:capacitance eq of state} there is also a electronic contribution $s_\text{cap}^2 = \frac{n^{(0)} e^2}{Cm}$ to the undamped speed of sound $s_0 \equiv \sqrt{s_\text{FL}^2 + s_\text{cap}^2}$.

\subsection{Parameter Estimates}
\label{sec:Parameter estimates}

To estimate parameters, we consider a graphene-hBN stack with gate-channel separation $d=100$ nm and average carrier density $n^{(0)} \sim 10^{12}$ cm$^{-2}$.
In graphene, the relaxation rate $\gamma \sim 650$ GHz and $\nu\sim 0.1$ m$^2/$s,\cite{Bandurin2016} so that the viscous length scale $r_d = \sqrt{\frac{\nu}{\gamma}} \sim 0.4\mu$m.\cite{Hui2020}
The relative dielectric constant of hBN is $\epsilon\sim 3.9$,\cite{Laturia2018,Torre2015} and we approximate $m$ and $e$ to be the bare electron mass and charge, respectively.
Therefore, the electronic contribution to sound is $s_\text{cap} \sim 0.9 \times 10^6$ m/s.
The speed of sound of Fermi liquids is $s_\text{FL} \sim v_F$,\cite{Landau1957} and Fermi velocities for metals are generally $v_F \sim 10^6$ m/s.\cite{AshcroftMerminBook}
Therefore, we will approximate the undamped speed of sound $s_0 \sim 2\times 10^6$ m/s.
Using the dispersion relation in Eq.~\ref{eq:longitudinal dispersion}, for $\omega = 1$ THz we have the true speed of sound $s \sim 1.9 \times 10^6$ m/s and attenuation coefficient $\alpha \sim 1/(6 \mu$m).
As a rough estimate, for characteristic lengths $h\sim L \sim 5 \mu$m the Reynolds numbers are $\operatorname{Re}_\nu \sim I/(160 \mu\text{A})$ and $\operatorname{Re}_\gamma \sim I/(26 \text{mA})$.
The ratio $\operatorname{Re}_\nu/\operatorname{Re}_\gamma \sim L^2/r_d^2$ is controlled by the viscous length scale $r_d \sim .4 \mu$m, so current micrometer-scale experiments will be in a regime where $\operatorname{Re}_\gamma$ tends to dominate the nonlinear behavior.
We remark that the apparent paradox that hydrodynamic effects could be dominated by momentum relaxation is due to linear-response considerations; by tuning the sample width $h$ such that $r_d\ll h$, a hydrodynamic description of the material remains valid but becomes indistinguishable from Ohm's law in the absence of convection.

\section{Electronic Bernoulli effect}

\begin{figure}
    \centering
    \includegraphics[width=.8\columnwidth]{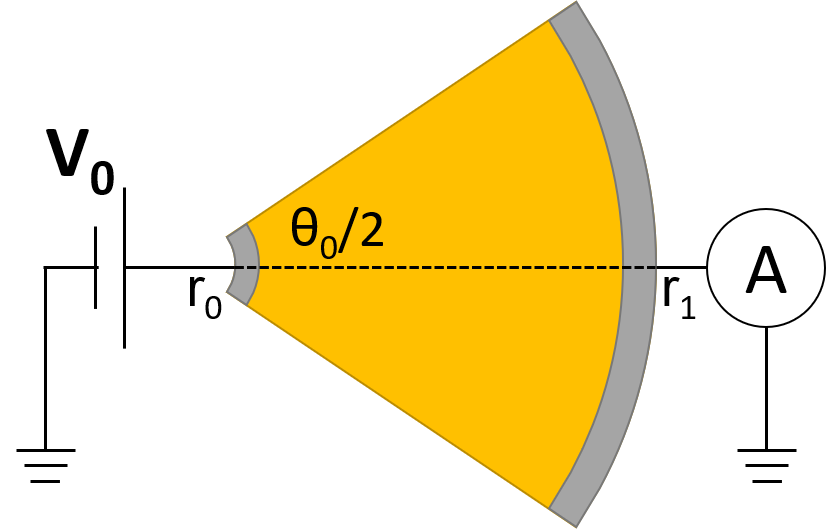}
    \caption{A topview of the Venturi geometry, with inner radius $r_0$ and outer radius $r_1$ and total wedge angle $\theta_0$.}
    \label{fig:venturi topview}
\end{figure}

\begin{figure}
    \centering
    \includegraphics[width=\columnwidth]{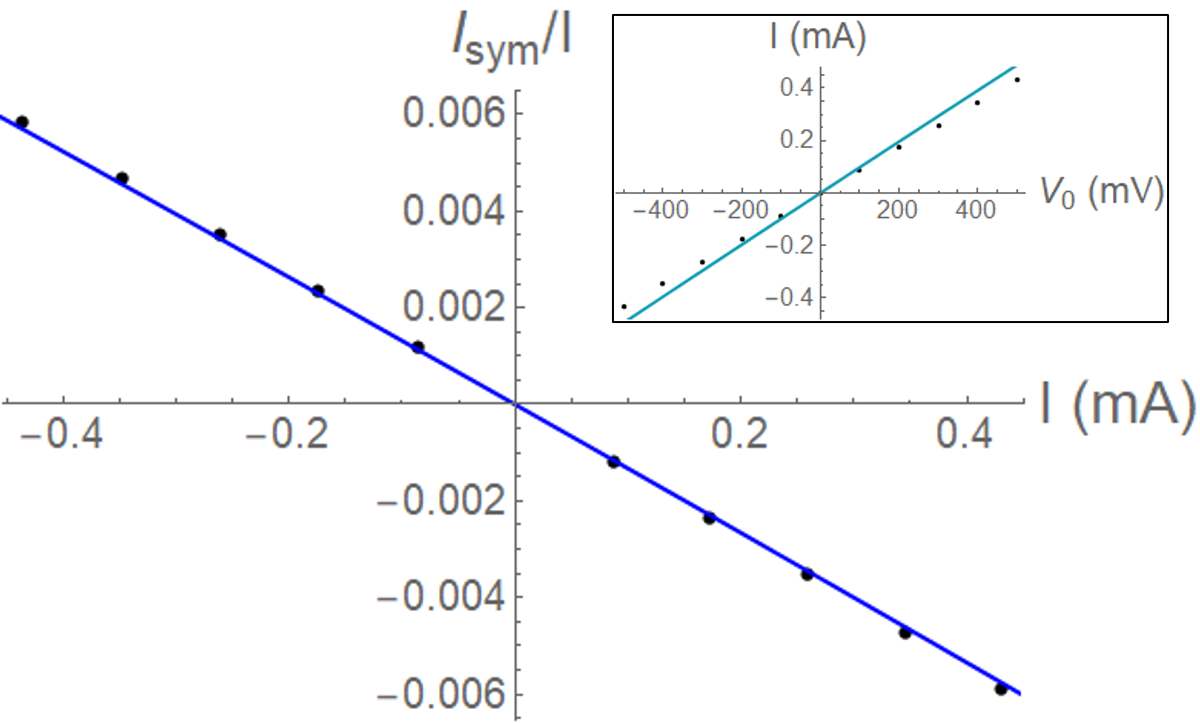}
    \caption{Main: A parametric plot of the voltage-symmetrized current $I_\text{sym}(V_0) \equiv \frac{1}{2}[I(V_0) + I(-V_0)]$ against total current $I(V_0)$. Inset: The I-V characteristic. The solid lines are obtained analytically from Eq.~\eqref{eq: Venturi Ohmic} in the $\nu \rightarrow 0$ limit, and the points are obtained numerically with finite $\nu$. Fixed-voltage boundary conditions are taken. The inner and outer radius are $5 \mu$m and $10 \mu$m respectively, with wedge angle $\theta_0 = \pi/2$, with graphene parameters $\nu = .1$ m$^2$/s and $\gamma = 650$ GHz. Since $r_d \sim .4 \mu$m and lengths are $\sim 10 \mu$m, viscous corrections to the analytic $\nu \rightarrow 0$ solution should be $\sim 5\%$.}
    \label{fig:venturi-numerics}
\end{figure}

We now apply the hydrodynamic formalism to derive a nonlinear contribution to the I-V characteristic $V\propto I^2$ in what we call the `Venturi' geometry (see Fig.~\ref{fig:venturi topview}), first analytically in the limit $\nu \rightarrow 0$.
For boundary conditions, we fix the voltage $\phi(r_0) = V_0$ and $\phi(r_1) = 0$ and take no-slip (vanishing velocity) at the side walls $\theta=\pm \theta_0/2$. We find that stationary purely radial "plug flow" ansatz $\mathbf{v} = v_r(r) \Theta(\theta_0^2-4\theta^2) \hat{\mathbf{r}}$ is a solution (with $\Theta$ the Heaviside step-function). Absence of viscosity is crucial as it allows for a zero-thickness boundary layer in this highly symmetric flow. 
The Navier-Stokes equation (Eq.~\eqref{eq:Navier-Stokes}) reduces to a simple ordinary differential equation
\begin{align}
    \frac{\partial}{\partial r}\left[e\phi + \frac{1}{2} m v_r^2\right] + m\gamma v_r = 0 \quad,
    \label{eq:Venturi Bernoulli}
\end{align}
where we have subsumed pressure into $\phi$ for simplicity.\footnote{When the device is gated, including $p$ is equivalent to renormalizing $e/m$. In the absence of gating, the long-range Coulomb interaction suppresses density fluctuations, so the pressure contribution is expected to be negligible.}
We further take the divergence-free (``incompressible flow'') ansatz $v_r = \frac{I}{\rho_e^{(0)} \theta_0} \frac{1}{r}$, where the yet-undetermined constant $I$ is the total current and $\rho_e^{(0)}$ is the average charge density.
Substituting this ansatz into Eq.~\eqref{eq:Venturi Bernoulli} and integrating from $r_0$ to $r_1$ (see Fig.~\ref{fig:venturi topview}), we obtain the nonlinear I-V characteristic
\begin{align}
    V_0 =&\frac{1}{\sigma_D} \left[\frac{l \ln(h_1/h_0)}{h_1 - h_0}I - \frac{1}{2}\left(\frac{1}{h_0^2} - \frac{1}{h_1^2}\right)\frac{I^2}{\rho_e \gamma}\right]
    \label{eq: Venturi Ohmic}
\end{align}
where $\sigma_D = \frac{n^{(0)} e^2}{m\gamma}$ is the Drude conductivity, $l = r_1 -r_0$ is the length, and $h_0 = \theta_0 r_0$ and $h_1 = \theta_0 r_1$ are the widths at the contacts.
The first term on the RHS corresponds to the Ohmic contribution,
while the second term is the nonlinear $I^2$ contribution from convection.
To further isolate the nonlinearity, we exploit the parity difference between the two contributions.
Because the nonlinearity is of even parity, a non-zero symmetrized current $I_\text{sym}(V_0) \equiv \frac{1}{2} [I(V_0) + I(-V_0)]$ provides a direct signature of the nonlinearity.
To estimate this effect, in Fig.~\ref{fig:venturi-numerics} we plot in blue the current fraction $I_\text{sym}/I$ and the I-V characteristic of Eq.~\eqref{eq: Venturi Ohmic} for wedge angle $\theta_0 = \pi/2$ with $r_0= 5\mu$m, $r_1=10 \mu$m, and graphene-hBN parameters as discussed in Sec.~\ref{sec:Parameter estimates}.
To incorporate a finite shear viscosity, which is difficult to solve analytically (see Appendix~\ref{sec: appendix Venturi}), we solve the Navier-Stokes equations numerically and plot the results as points in Fig.~\ref{fig:venturi experiment}.
The exact ($\nu=0$) result of Eq.~\eqref{eq: Venturi Ohmic} matches well with the numerical result, as expected of the fact that the viscous length scale $r_d \equiv \sqrt{\frac{\nu}{\gamma}} \ll r_0\theta_0$ is small for experimentally relevant parameters.
As demonstrated by Fig.~\ref{fig:venturi experiment}, this nonlinear effect ($I_\text{sym} \sim 400$ nA for $I\sim 200 \mu$A) should be experimentally measurable.

This nonlinear I-V characteristic in electronic hydrodynamics is the analogue of the Bernoulli effect in classical hydrodynamics, the prototypical example of convective acceleration, which is traditionally demonstrated using a Venturi tube.
The Bernoulli effect is typically demonstrated in an inviscid fluid of divergence-free (incompressible) flow, analogous to our assumptions.
In fact, the classical Bernoulli (energy conservation) equation is analogous to Eq.~\eqref{eq:Venturi Bernoulli}; the term in brackets corresponds to the classical Bernoulli contribution, while the $\gamma$ term accounts for the additional dissipation from a finite conductivity.
As a result, the nonlinear term of the I-V characteristic Eq.~\eqref{eq: Venturi Ohmic} can be calculated exactly by classical Bernoulli considerations.

We turn to the subtle issue of solving for the total current $I(V_0)$ given the input voltage $V_0$, i.e. verifying that the ansatz satisfies the boundary conditions.
Because this requires solving a quadratic equation for $I$, the solution is generically multivalued and may not even have a solution.
In the limit of small $V_0$, linear response must provide the correct answer on physical grounds; this selects the solution branch continuously connected to the solution $I=0$ at $V_0=0$, where parity was broken by $\gamma$.
The opposite branch is therefore expected to be unstable to $\theta$-dependent perturbations.
The region where the purely radial solution does not exist corresponds to particle flow in the divergent direction; for classical fluids, it is known that divergent flow eventually becomes unstable and develops turbulence.\cite{LandauVol6,RosenheadBook}
To estimate the scale of nonlinearity at which the radial ansatz fails, one can define a Reynolds number
\begin{align}
    \operatorname{Re}_\gamma \equiv& \frac{\int_{r_0}^{r_1} dr F_{\text{conv},r}}{-\int_{r_0}^{r_1} dr \rho\gamma v_r} = \frac{-1}{2lh_0}\frac{I}{\rho_e \gamma} \left[\frac{\frac{h_1}{h_0} - 1}{ \ln \frac{h_1}{h_0}} \left(1 -\frac{h_0^2}{h_1^2}\right) \right]
\end{align}
which is precisely the ratio of the two terms in Eq.~\eqref{eq: Venturi Ohmic}.
The instability point occurs at $\operatorname{Re}_\gamma = -1/2$.
We summarize the resolution of these subtleties in Fig.~\ref{fig:venturi experiment}.

Finally, we now highlight three aspects of the Bernoulli non-linearity that should help identify it unambiguously in experiments. 
To start, following Eq. \eqref{eq: Venturi Ohmic} we note that the quadratic term is independent of the momentum relaxation parameter $\gamma$, and hence may be identified by comparing I-V traces taken at different temperatures or even from different samples of the same material. 
Secondly, the simple charge density-dependence may be probed by varying backgate voltage.
After factoring out the density-dependent Drude resisitivity $1/\sigma_D$ (cf. Eq.~\ref{eq: Venturi Ohmic}), the nonlinear term only has an inverse dependence on charge density (and its sign depends on the carrier charge).
Lastly, Eq.~\eqref{eq: Venturi Ohmic} has a distinct geometric dependence interpolating in a somewhat unusual way between conventional and ballistic transport. 
For a fixed aspect ratios $h_1/h_0$ and $l/h_0$, we find that the Ohmic resistance contribution scales with the size of the device as $1/h_0$ while the nonlinear Bernoulli contribution scales as $1/h_0^2$.
In addition, the Ohmic resistance contribution has the conventional linear scaling with length $l$, while the nonlinear Bernoulli contribution has the $l$-independent hallmark of $\emph{ballistic}$ transport.

\section{Eckart Streaming: A ``Hydrodynamic Solar Cell''}
\label{sec: Eckart}

A dramatic effect of nonlinearity occurs upon applying an oscillatory drive: down-conversion.
In a backgated device of length $l$ and width $h$ (see Fig.~\ref{fig:eckart experiment}), we consider setting up a traveling longitudinal (sound) wave by application of a voltage oscillation $\phi(x=0) = V_0 \cos\omega t$ at the left contact with the right contact grounded ($\phi(x=l)=0$).
This will result in a DC current via the down-conversion sourced by the convective force (Eq.~\eqref{eq:F_conv}).
Such a device can be described as a ``hydrodynamic solar cell'' providing a DC photocurrent if the (localized) voltage oscillation is driven by EM radiation.
For simplicity, we will focus on bulk dissipation (i.e. attenuation due to $\alpha>0$) contributions to the convective force and neglect those of boundary dissipation, which only results in a quantitative underestimate of the DC current (see Appendix~\ref{sec: appendix Eckart boundary dissipation}).
This is the electronic analogue of Eckart streaming in classical hydrodynamics, where the convective force is primarily generated by bulk dissipation.\cite{Eckart1948,Nyborg1965,Lighthill1978,Wiklund2012}
To see this, we need to solve the full Navier-Stokes equation (Eq.~\eqref{eq:Navier-Stokes}), whose nonlinearity precludes a single-mode ansatz.
To handle this, we will seek a perturbative solution in the input voltage amplitude $V_0$ (see Appendix~\ref{sec: appendix Eckart} for full mathematical detail).

\subsection{Perturbative Calculation}

We begin by expanding the hydrodynamic variables in a power series expansion of $V_0$, e.g. $\rho = \rho^{(0)} + \rho^{(1)} + \rho^{(2)} + \ldots$; $\rho^{(0)}$ corresponds to the equilibrium mass density, while $\rho^{(1)}$ and $\rho^{(2)}$ are the first and second order solutions.
At leading (linear) order, the single-mode ansatz $\phi^{(1)} \sim V_0 e^{i(\pm k_l x - \omega t)}$ along $x$ with wavenumber $k_l = k+i\alpha$ is appropriate.
Imposing the fixed-voltage boundary conditions, the solution of $\phi^{(1)}$ is a traveling wave with
reflected component; the grounded edge acts as a mirror.
Because of the backgate providing a capacitance per area $C$, the voltage oscillation of amplitude $V_0$ sets up a charge density oscillation $\rho_e^{(1)} = C\phi^{(1)}$ of amplitude $CV_0$ (see Eq.~\eqref{eq:capacitance eq of state}).
Via the density continuity equation (Eq.~\eqref{eq:density continuity}), the density oscillations drive a longitudinal velocity oscillation $v_x^{(1)}$, schematically written as
\begin{align}
    v_x^{(1)} \sim u_0\Re\left[e^{(ik-\alpha)x-i\omega t} + e^{(ik-\alpha) (2l-x)-i\omega t}\right]
\end{align}
where $\Re$ denotes real part and $u_0 = \frac{CV_0}{\rho_e^{(0)}} \frac{\omega}{|k_l|}$ is the velocity amplitude.
We also take a no-slip boundary condition, which is \emph{not} satisfied by $v_x^{(1)}$.
However, as previously stated we will neglect the boundary corrections to $v_x^{(1)}$ for simplicity (see Appendix~\ref{sec: appendix Eckart boundary dissipation}).
As a result, the leading order solution $v_x^{(1)}$ results in a DC convective force (see Eq.~\eqref{eq:F_conv})
\begin{align}
    \overline{F^{(2)}_{\text{conv},x}} =& \rho^{(0)}u_0^2\frac{\alpha \sinh[2\alpha(l-x)] - k\sin[2k(l-x)]}{\cosh 2\alpha l -\cos 2kl}
    \label{eq:Eckart F_conv}
\end{align}
where the overbar denotes time-average.
The first term in the numerator arises from the bulk dissipation $\alpha$, while the second term arises from interference effects; in the limit $\alpha l \gg 1$, where interference effects are small, the RHS of Eq.~\eqref{eq:Eckart F_conv} simplifies to $\alpha e^{-2\alpha x}$.
This rectified DC force will result in a DC current.

\begin{figure}
    \centering
    \includegraphics[width=\columnwidth]{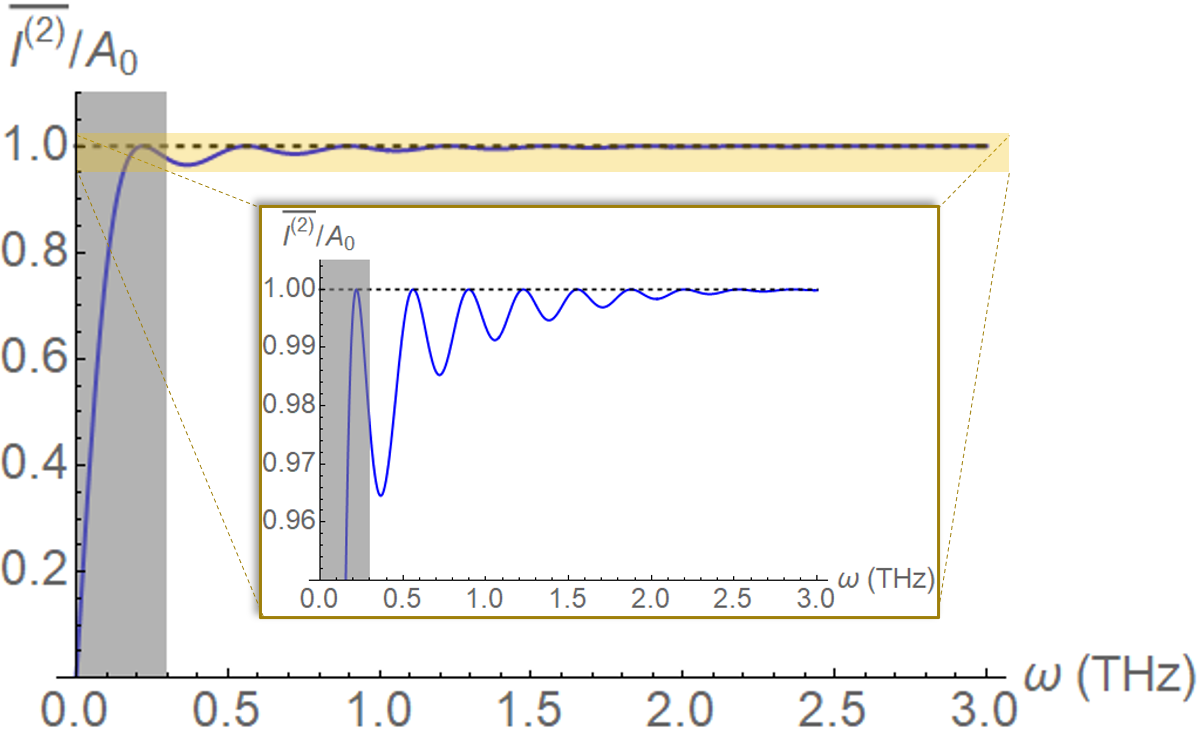}
    \caption{Main: A plot of $\overline{I^{(2)}}$ at fixed input current amplitude $I_0$ for device length $l=30 \mu$m and graphene-hBN parameters stated in Sec.~\ref{sec:Parameter estimates}, in units of $A_0 = \frac{I_0^2}{\rho_e^{(0)} h} \frac{1}{2l\gamma} \left(1-\frac{2r_d}{h}\tanh\frac{h}{2r_d}\right)$. We remark that this is also a scaled plot of the Reynolds number $\operatorname{Re}_\gamma$. Inset: A blowup of the yellow highlighted portion. At high frequencies, $\operatorname{Re}_\gamma$ saturates to a constant $A_0 = \frac{I_0}{\rho_e^{(0)}h}\frac{1}{2l\gamma}$, while at sufficiently low frequencies the interference oscillations become more visible. The gray box demarcates the low frequency region $\omega \ll \gamma$, where perturbation theory in $V_0$ breaks down for a fixed $I_0$.}
    \label{fig:Eckart frequency}
\end{figure}

We now solve for the second-order DC current $\overline{I^{(2)}}$.
The DC current density $\overline{\mathbf{J}^{(2)}} \equiv \rho_e^{(0)} \overline{\mathbf{v}^{(2)}} + \overline{\rho_e^{(1)} \mathbf{v}^{(1)}}$ must be divergence-free to satisfy current conservation (i.e. density continuity Eq.~\eqref{eq:density continuity}).
With the ansatz $\overline{v_y^{(2)}} = 0$, this implies that the current density $\overline{\mathbf{J}^{(2)}} = \overline{J_x^{(2)}}(y) \mathbf{\hat{x}}$ only varies along $y$.
However, the convective force given by Eq.~\eqref{eq:Eckart F_conv} varies along $x$.
This paradox is resolved by static screening, where the $x$-dependence of convection will be canceled by contributions from the effective voltage $\overline{\phi_\text{eff}^{(2)}} \equiv \overline{\phi^{(2)}} + \frac{1}{\rho_e^{(0)}} \overline{p^{(2)}}$.
Utilizing separation of variables in the NS equation (Eq.~\eqref{eq:Navier-Stokes}), we can solve for $\overline{\phi_\text{eff}^{(2)}}$ by applying the voltage-fixed boundary conditions $\overline{\phi^{(2)}}(x=0) = \overline{\phi^{(2)}}(x=l) = 0$.
Therefore, the ``screened'' convective force (which is no longer spatially dependent) becomes
\begin{align}
    \overline{F_{\text{conv},x}^{(2)}} - \rho_e^{(0)}\frac{\partial \overline{\phi_\text{eff}^{(2)}}}{\partial x} =& \frac{1}{l}\int_0^l dx \overline{F_\text{conv}^{(2)}}\ .
    \label{eq:Eckart Poiseuille}
\end{align}
Solving NS for the current density $\overline{J_x^{(2)}}$ and integrating across the channel to get the total current $I^{(2)}$, we get
\begin{align}
    \overline{I^{(2)}} =& \frac{I_0^2}{\rho_e^{(0)} h}\frac{1}{2l\gamma}\left[1 - \frac{2 - 2\cos 2kl}{\cosh 2\alpha l - \cos 2kl}\right] 
    \nonumber
    \\
    &\phantom{\frac{I_0^2}{\rho_e^{(0)} h}\frac{1}{2l\gamma}} \times \left(1 - \frac{2r_d}{h} \tanh \frac{h}{2r_d} \right)
    \label{eq:Eckart current}
\end{align}
where $I_0 \equiv \rho_e^{(0)} h u_0$ is the input current amplitude and have assumed that convection provides the dominant DC force (see Appendix~\ref{sec: appendix Eckart forces}).
The term in parentheses is a viscous correction, reflecting the $y$-dependence of the current flow due to no-slip.
The bracketed terms correspond to dissipation and interference contributions from the convective force (Eq.~\eqref{eq:F_conv}), respectively.
The effect of these contributions is demonstrated in Fig.~\ref{fig:eckart experiment}, where we have schematically plotted the dependence of DC current on the channel length $l$.
In the limit $\alpha l \ll 1$, the interference term dominates, leading to oscillatory behavior controlled by $kl$.
In the opposite limit $\alpha l \gg 1$, the interference term becomes negligible, and the DC current scales as $\overline{I^{(2)}}\sim l^{-1}$.
Other than the device length $l$, one could also study the frequency dependence of Eq.~\eqref{eq:Eckart current} (via $k_l(\omega) = k+i\alpha$), which is plotted in Fig.~\ref{fig:Eckart frequency} for a fixed $I_0$.\footnote{Because $I_0 = \rho_e^{(0)} h \frac{CV_0}{\rho^{(0)}_e} \frac{\omega}{|k_l|}$, perturbation theory will break down for sufficiently low $\omega$. This happens when $\omega \ll \gamma$, i.e. when $k_l$ is dominated by $\gamma$ and tends to a constant.}
Similarly, interference effects appear at low frequencies and become negligible at high frequencies.

\subsection{Discussion and Estimates}

An effect similar to Eckart streaming was previously discussed by Dyakonov and Shur\cite{Dyakonov1996} and extended in Ref.~\onlinecite{Torre2015}.
They envisaged operating with zero DC current bias $\overline{I} = 0$ instead of zero DC voltage drop, so that one generates a DC voltage instead of a DC current.
These theoretical treatments\cite{Dyakonov1996, Torre2015} similarly neglected boundary dissipation, which only leads to quantitative corrections to DC voltage.
However, for their case boundary dissipation leads to qualitative flow corrections (see Appendix~\ref{sec: appendix Eckart boundary dissipation}); further discussion is deferred to Sec.~\ref{sec: Rayleigh}. 
We point out that, in either case, if the voltage oscillation is driven by an impingent EM wave, the device is a ``hydrodynamic solar cell'' generating a DC photocurrent (photovoltage).
In contrast to typical solar cells (e.g. a p-n junction), the hydrodynamic solar cell does not break parity by construction; parity is intrinsically broken by dissipation, setting the direction of the photocurrent.
Therefore, Eckart streaming provides a novel mechanism for photocurrent (photovoltage) generation. Signatures of downconverted DC voltage generation by THz radiation have been measured in ultraclean 2DEGs.\cite{Tauk2006, Vicarelli2012,Giliberti2015,Bandurin2018}

One can define Reynolds numbers to estimate the strength $I^{(2)}/I_0$ of the nonlinearity.
The Reynolds number $\operatorname{Re}_\gamma$ for this system can be defined as
\begin{align}
    \operatorname{Re}_\gamma \equiv& \frac{\frac{1}{l}\int_0^l \overline{F_{\text{conv},x}^{(2)}}}{\rho_e^{(0)}\gamma u_0}= \frac{I_0}{\rho_e^{(0)} h}\frac{1}{2l\gamma}\left[1 - \frac{2 - 2\cos 2kl}{\cosh 2\alpha l - \cos 2kl}\right]
    \label{eq:Eckart Reynolds number}
\end{align}
which explicitly appears in Eq.~\eqref{eq:Eckart current}.
The viscous Reynolds number can be similarly defined such that $\operatorname{Re}_\nu = \frac{h^2}{r_d^2} \operatorname{Re}_\gamma$, where we approximate the viscous gradients to have length scale $L=h$ (see Eq.~\eqref{eq:Re_nu}).
The contribution from $\operatorname{Re}_\nu$ is hidden within $r_d$; in the limit $r_d \gg h$ where viscous contributions dominate, $\operatorname{Re}_\nu$ can be made manifest by perturbatively expanding Eq.~\eqref{eq:Eckart current} in $h/r_d$. 
Since $r_d\gg h$ for the experimental systems of interest, the Reynolds number $\operatorname{Re}_\gamma \sim I^{(2)}/I_0$ corresponds to the scale of DC current (up to a small viscous correction).

We now estimate the size the DC current in experiment (see Appendix~\ref{sec: appendix hydro modes} for dispersion relations).
We take device size $l=50 \mu$m and $h=5\mu$m and operate at $\omega = 1$ THz, with graphene-hBN parameters from Sec.~\ref{sec:Parameter estimates}; for these choices, the interference effects are small since $\alpha l \sim 5$.
Therefore, we find $\operatorname{Re}_\gamma \sim I_0/(312 \text{mA})$ and therefore $\overline{I^{(2)}}/\text{nA} \sim (I_0/24\mu\text{A})^2$.
Observing the oscillatory effects is more difficult, requiring smaller $l$ and more measurement precision.
Despite this, in an optimistically sized device of length $l=20\mu$m, we plot the frequency dependence of $\operatorname{Re}_\gamma$ in Fig.~\ref{fig:Eckart frequency}. 
The oscillations are suppressed by a factor of $0.01$; if one asks for a streaming current $\overline{ I^{(2)}} \sim 1$ nA, the oscillations will be of order 10 pA.
We therefore conclude that an Eckart streaming current should be visible in current experiments, with interference oscillations being a challenging observable.

\section{Rayleigh Streaming}
\label{sec: Rayleigh}

We now turn to the limit where boundary dissipation dominates, i.e. the bulk dissipation $\alpha$ is negligible.
Here, the no-slip condition is critical.
In a rectangular backgated device of width $h$ (see Fig.~\ref{fig:rayleigh experiment}), we consider setting up a longitudinal standing wave of wavelength $\lambda \gg \alpha^{-1}$ along $x$.
In this case, the system cannot support a finite DC current due to reflection symmetry in $y$.
Therefore, down-converted DC current flows sourced by the convective force (see Eq.~\eqref{eq:F_conv}) must circulate.
The circulating current leads to a measurable orbital magnetization of wavelength $\lambda/2$ along $x$ with reflection-symmetric modulation along $y$ (see Fig.~\ref{fig:rayleigh experiment}).
This is the analogue of Rayleigh streaming in classical hydrodynamics, where the convective force is primarily generated by boundary dissipation.\cite{Lighthill1978, Nyborg1965, Riley2001}
Remarkably, localized boundary effects lead to nontrivial flows throughout the bulk (see Appendix~\ref{sec: appendix Rayleigh} for full mathematical detail).

\subsection{Perturbative Calculation}

We begin by working perturbatively in the input current amplitude $u_0$, where at linear order we take the longitudinal wave ansatz
\begin{align}
    v_{l,x}^{(1)}= u_0 \sin kx \cos \omega t
    \label{eq:Rayleigh ansatz}
\end{align}
This is consistent with a current-fixed boundary condition $J_x(x=0)=0$ (i.e. DC current $\overline{I} = 0$).
For simplicity, we work in a semi-infinite strip of width $h$ (i.e. $|y| \leq h/2$ and $x\geq 0$) with the above current-fixed boundary condition.
To satisfy no-slip, a transverse mode $\mathbf{v}_t^{(1)}$ is necessary to correct the total flow $\mathbf{v}^{(1)} = \mathbf{v}_l^{(1)} + \mathbf{v}_t^{(1)}$.
This transverse correction disperses along $y$ with wavenumber $k_t = k_t' + ik_t''$, and hence forms a ``boundary layer'' of size $1/k_t''$ exponentially localized to the wall.
We will work in the thin boundary layer and long wavelength limit $k_t''^{-1} \ll h \ll \lambda$.
In this limit, the resulting convective force (see Eq.~\eqref{eq:F_conv}) can be schematically written as
\begin{align}
    \overline{F_{\text{conv},x}^{(2)}} \sim \rho^{(0)} u_0^2 \, k \, e^{-k_t'' y_+} \sin 2kx + (y\leftrightarrow -y)
\end{align}
where $y_+ = y+\frac{h}{2}$ is the distance from the lower boundary.\footnote{More precisely, this is schematic form of the ``screened'' convective force with contributions from the effective voltage $\phi_\text{eff}^{(2)} = \phi^{(2)} + \frac{1}{\rho^{(0)}}p^{(2)}$.}
As a result of the quadratic non-linearity, the wavelength of the convective force is halved to $\lambda/2$.
In addition, the convective force is localized to the boundary layer, reflecting the fact that convection is driven by boundary dissipation.
It is therefore convenient to divide the flow into bulk and boundary-layer regions, stitched together at the interface.
Despite the localized nature of the convective force, its effect will persist into the bulk by providing a slip boundary condition.

Now, we study the second-order DC flow.
We first consider the boundary layer region, assuming that the viscous length scale $r_d\equiv \frac{\nu}{\gamma}\ll h$.
The convective force localized to the boundary layer of size $1/k_t''$ leads to a localized flow along $x$.
Because of the shear viscosity $\nu$, the boundary layer momentum propagates into the bulk with the viscous length scale $r_d$.
Therefore, the boundary layer ``screens'' the no-slip condition, providing instead a slip velocity for the bulk flow.
This slip velocity can be written as $v_\text{slip}^{(2)} \sin 2kx$, where schematically $v_\text{slip}^{(2)} \sim \frac{u_0^2 k}{4 \gamma} e^{-1/k_t'' r_d}$.
Equipped with the slip boundary, we now solve the NS equation (Eq.~\eqref{eq:Navier-Stokes}) for the bulk flow where the convective force vanishes and obtain
\begin{align}
    \overline{J^{(2)}_{\text{bulk},x}} =& J_\text{slip}^{(2)}\sin 2kx \left[ \frac{\frac{h}{2r_d}\cosh \frac{y}{r_d} - \sinh\frac{h}{2r_d}}{\frac{h}{2r_d}\cosh \frac{h}{2r_d} - \sinh\frac{h}{2r_d}}\right]
    \label{eq:Rayleigh vx}
    \\
    \overline{J^{(2)}_{\text{bulk},y}} =& J_\text{slip}^{(2)}2k r_d\cos 2kx \left[ \frac{- \frac{h}{2r_d} \sinh \frac{y}{2r_d} + \frac{y}{r_d}\sinh\frac{h}{2r_d}}{\frac{h}{2r_d}\cosh \frac{h}{2r_d} - \sinh\frac{h}{2r_d}}\right]
    \label{eq:Rayleigh vy}
\end{align}
The slip current $J_\text{slip}^{(2)} \equiv \rho_e^{(0)} v_\text{slip}^{(2)}$ results from boundary convection, while the term in brackets is a geometric factor resulting from satisfying the slip velocity boundary condition.
The DC current flow is plotted in Fig.~\ref{fig:rayleigh experiment}, where it is clear that the current circulates in cells of length $\lambda/4$ and width $h/2$.

\subsection{Discussion and Estimates}

A previous related proposal by Dyakonov and Shur\cite{Dyakonov1996} and its recent extension\cite{Torre2015} discussed downconversion effects with a current-fixed boundary $J(x=0) = 0$, similar to this case. 
However, they instead took a stress-free boundary condition which has no boundary dissipation.
In their case, there is no circulating current; without boundary-layer contributions, the convective force only leads to an excess of DC voltage (see Appendix~\ref{sec: appendix Rayleigh}).
Therefore, Rayleigh streaming is qualitatively distinct from previous nonlinear proposals in electron hydrodynamics.

Since the effect of the convective force is to generate a slip velocity $v_\text{slip}^{(2)}$, we can estimate the scale $v_\text{slip}^{(2)}/u_0$ by an appropriate Reynolds number.
The Reynolds number $\operatorname{Re}_\gamma$ is defined in this case to be
\begin{align}
    \operatorname{Re}_\gamma \equiv \frac{\max \overline{F_\text{conv}^{(2)}}}{\rho^{(0)}\gamma u_0} = \frac{I_0}{\rho_e^{(0)} h}\frac{k}{4\gamma} f(\omega/\gamma)
    \label{eq:Rayleigh Reynolds number}
\end{align}
where $f$ is a dimensionless function of $\omega/\gamma$ described in Appendix~\ref{sec: appendix Rayleigh}.\footnote{As before, we only consider the ``screened'' convective force in the above equation, equivalent to including boundary contributions only.}
We remark that $f$ develops an interesting resonance at $\omega = \frac{\sqrt{5}}{2}\gamma$ where perturbation theory breaks down, but we operate away from this point and will not discuss it further.
It turns out $\operatorname{Re}_\gamma e^{-1/k_t''r_d} = v_\text{slip}^{(2)}/u_0$, i.e. slip velocity is given by the Reynolds number up to an exponential factor controlled by the viscous length scale $r_d$.
However, the viscous Reynolds number $\operatorname{Re}_\nu$ does not contribute to the effect; in the limit $\gamma \rightarrow 0$, the scale $v_\text{slip}^{(2)}/u_0$ is instead set by the Mach number $u_0 k/\omega$.
Despite the necessity of a finite shear viscosity $\nu$ to generate a convective force, $\operatorname{Re}_\nu$ does not set the scale $v_\text{slip}^{(2)}$ of the result; this curious fact was first remarked by Rayleigh\cite{Rayleigh1883} (see Appendix~\ref{sec: appendix Rayleigh} for additional discussion).

\begin{figure}
    \centering
    \includegraphics[width=\columnwidth]{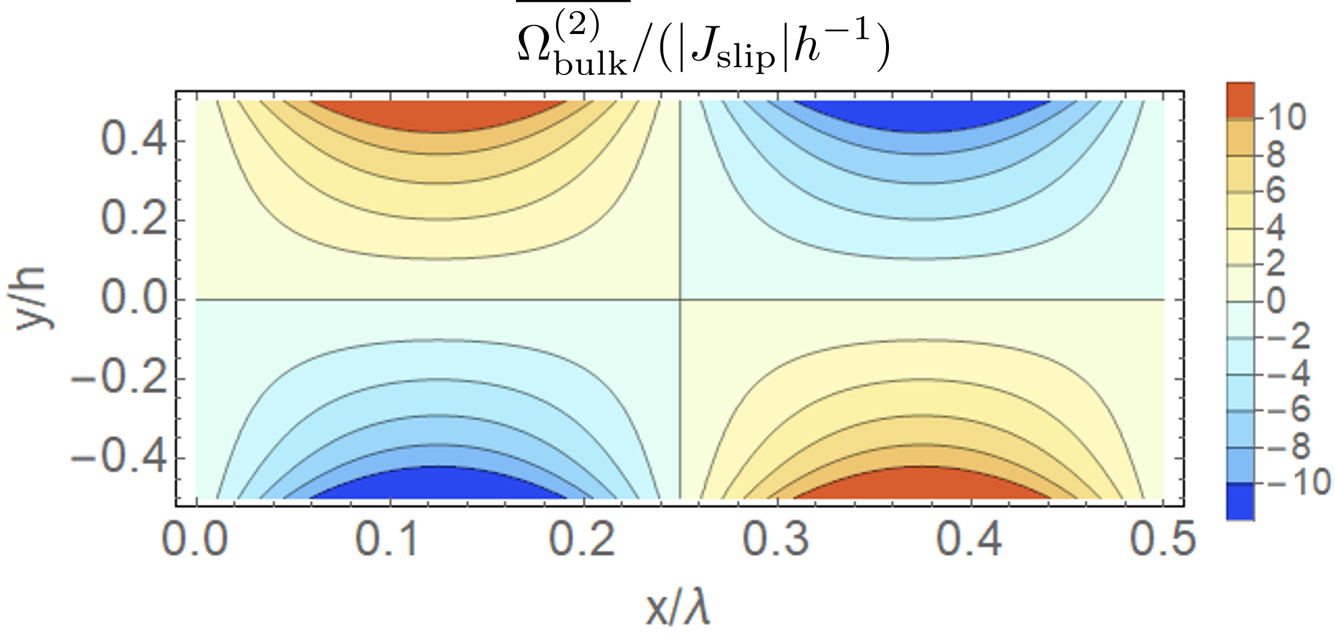}
    \caption{A plot of the bulk vorticity distribution $\overline{\Omega_\text{bulk}^{(2)}} \equiv \nabla \times \overline{\mathbf{J}_\text{bulk}^{(2)}}$ induced by Rayleigh streaming for $h=5 \mu$m and $\omega = 2$ THz with graphene-hBN parameters as in Sec.~\ref{sec:Parameter estimates}. The local bulk vorticity corresponds to a Coulomb-like point source of magnetic field due to Ampere's law.}
    \label{fig:rayleigh magnetization}
\end{figure}

We propose that the circulating flow profile could be detected via magnetometry.
To estimate the effect in realistic systems, we set $\omega = 2$ THz and channel width $h = 5\mu$m with graphene-hBN parameters as in Sec.\ref{sec:Parameter estimates} (see Appendix~\ref{sec: appendix hydro modes} for dispersion relations).
We first verify the assumptions we made: $k_t''^{-1} \ll h \ll \lambda$, $r_d \ll h$, and $\alpha \ll k$.
These are $k_t'' h \sim 13$, $h/\lambda \sim 0.80$, $r_d/h \sim .08$ and $\alpha/k \sim 0.2$, so we expect our solution to be roughly correct.
For the scale of the DC effect, we find $\operatorname{Re}_\gamma \sim I_0/(23 \text{mA})$ and $k_t'' r_d \sim 1.1$, so that $v_\text{slip} \sim (I_0/71 \text{mA}) u_0$.
Since Ampere's law implies $-\nabla^2 B_z = \mu_0 \nabla \times \mathbf{J} \delta (z)$, the vorticity $\Omega \equiv \nabla \times \mathbf{J}$ acts as a Coulomb-like point source of magnetic field.
The vorticity is plotted for these parameters in Fig.~\ref{fig:rayleigh magnetization}, where it is concetrated near the edges since the viscous length scale $r_d \ll h$ is small.
To make a rough estimate of the magnetic field strength, we take $\overline{B_z} \sim \frac{\mu_0}{z}\int_\text{cell} \nabla \times \overline{\Omega_\text{bulk}^{(2)}}$ at a height $z$ from the sample; we approximate the magnetic field to be sourced by the net circulation in the nearest vortical cell.
This gives $B_z \sim \frac{(I_0/9.3 \mu\text{A})^2}{z/\mu\text{m}} \times 10^{-10} \text{T}$.
Therefore, the magnetic fields should be detectable for $I_0\sim 9.3 \mu$A by scanning SQUID magnetometers.

\section{Summary and Outlook}

This paper argues for using non-linear DC transport and other manifestations of convective nonlinearity to identify and study electron hydrodynamics.
We have laid out three electronic analogues of nonlinear classical phenomena - the Bernoulli effect, Eckart streaming, and Rayleigh streaming - which lead to an experimentally measurable nonlinear I-V characteristic, down-converted DC current, and DC current vortices, respectively (see Fig.~\ref{fig:1}).
We have opted to derive and discuss all three effects using the familiar Navier-Stokes formalism, leaving a more complete microscopic treatment for future work. All three effects result from the interplay of the non-dissipative and nonlinear convection force with other dissipative contributions in Navier-Stokes from viscosity and momentum relaxation. 

It is interesting to note that interactions do not play an explicit role in our results -- both convection and momentum relaxation (the dominant form of relaxation) are well understood in the non-interacting limit of the many-electron problem. 
Instead, strong electron-electron interactions justify the coarse-grained effective description, removing the need to consider the complications of quasi-particle physics.
In particular, local equilibration (assumed throughout) is likely to be violated in the limit of weak interactions, requiring a more systematic microscopic treatment. 
This will be required, for example, before extrapolating our results to low temperatures. 

To obtain stronger nonlinear signatures, one would like to make the Reynolds numbers $\operatorname{Re}_\nu$ and $\operatorname{Re}_\gamma$ as large as possible.
Since the viscous length scale $r_d^2 = \nu/\gamma$ is typically smaller than the characteristic lengths in experiment, $\operatorname{Re}_\gamma$ is the limiting factor. 
In addition to reducing the momentum relaxation rate $\gamma$, one could also reduce the density $n$ at fixed current to improve the Reynolds numbers; particles must move more rapidly to maintain the current.
Therefore, nonlinear effects should be most prominent in clean, low-density hydrodynamic materials.
Our focus has been away from linear response, which is a bedrock foundation of experimental condensed matter physics. 
Nonlinear phenomena are comparatively more difficult to interpret and tend to be less explored, especially with the purpose of extracting basic information, e.g. where in the phase diagram a given material happens to be. 
However, since our primary focus has been on \emph{leading} deviations from linear response, we are nonetheless optimistic that identifying electron hydrodynamics from nonlinear behavior is feasible. 

In particular, the detection of the AC-generated static current described above would provide strong evidence for the presence of hydrodynamic behavior. 
Additionally, hydrodynamic nonlinearities should also generate upconverted 2$f$ signals, which we leave to future work.
This also tantalizingly suggests the possible utility of hydrodynamic materials as a novel platform for creating nonlinear electronic devices.\cite{Dyakonov1993,Dyakonov1996}
The nonlinear I-V characteristic of the Venturi wedge device clearly displays the onset of instability phenomena far separated from linear response. Such convective instabilities are a known route to classical turbulence\cite{LandauVol6,RosenheadBook}, i.e. in the absence of momentum relaxation. 
In the electronic system, where momentum relaxation dominates and viscous length scale $r_d$ is short, we suspect that the behavior may be qualitatively distinct from turbulence.\footnote{Hui, Oganeysan, and Kim. Work in progress.}
These and other phenomena pose a fertile frontier for near-term exploration of electron hydrodynamics.

\vspace{5mm}
{\noindent\bf Acknowledgements}
We thank Brad Ramshaw, Kin Fai Mak, Jie Shan, Peter Armitage, Sriram Ganeshan, Alexander Abanov, Minwoo Jung, and Maytee Chantharayukhonthorn for helpful discussions.

\bibliography{biblio}

\appendix
\onecolumngrid

\section{Oscillatory Hydrodynamic Modes}
\label{sec: appendix hydro modes}
Here we study the hydrodynamic modes at linear order (without boundary conditions), where the convective term $\mathbf{F}_\text{conv}$ is neglected.
Because of linearity, the harmonic modes will not mix; the linear-order ansatz $\mathbf{v}^{(1)} \propto e^{i(kx - \omega t)}$ is appropriate. 
We eliminate the variables $p$ and $\phi$ in Navier-Stokes (Eq.~\eqref{eq:Navier-Stokes}) by using density continuity (Eq.~\eqref{eq:density continuity}) as well as the equations of state (Eq.~\eqref{eq:pressure eq of state} and Eq.~\eqref{eq:capacitance eq of state}).
The resulting dispersion relation can be separated in longitudinal ($\nabla\times\mathbf{v}^{(1)} = 0$) and transverse ($\nabla\cdot\mathbf{v}^{(1)}=0$) contributions, and are given by
\begin{align}
    \omega_l^2 =& \left(s_0^2 - i\omega_l \left[2\nu+ \tilde{\zeta}\right]\right)k_l^2  - i\omega_l \gamma \label{eq:longitudinal dispersion}\\
    \omega_t =& i\nu k_t^2 - i\gamma \label{eq:transverse dispersion}
\end{align}
where $s_0^2 = s_\text{FL}^2 + s_\text{cap}^2$.
The longitudinal dispersion describes a damped sound wave with undamped speed $s_0$; both pressure and electric forces contribute additively to $s_0$ as a result of the equations of state.
In particular, the electronic contribution relies on backgate screening of the Coulomb interaction to achieve this form.
The transverse dispersion describes the propagation of incompressible shear oscillations, whose spatial extent is controlled by the viscous length scale $r_d$; a finite shear viscosity is necessary for the transfer of momentum into adjacent layers.
In contrast to the longitudinal case, the transverse modes do not drive density oscillations and therefore do not generate pressure or electric forces.
Therefore, the transverse result is independent of the equations of state, and in particular it does not depend on the presence of a backgate.

We remark that measuring the attenuation of longitudinal and transverse oscillations would provide direct, boundary-independent measures of both shear and bulk viscosity, as opposed to DC flow profiles which require the boundary\cite{Moll2016, Sulpizio2019, Ku2019} or inhomogenous current injection profiles\cite{Bandurin2016, Levin2018} to enforce velocity gradients.
A careful experimental study of finite-frequency behavior of hydrodynamic materials has yet to be done even at linear order, as far as the authors are aware; in particular, this could provide new cross-checks of previous viscosity measurements.
A proposal for for a shear viscometer utilizing oscillatory motion was made in Ref.~\onlinecite{Tomadin2014}.

\section{Electronic Venturi Effect - Treating Viscosity}
\label{sec: appendix Venturi}
The full problem, with both finite (kinematic) shear viscosity $\nu$ and momentum relaxation $\gamma$ is challenging.
Because viscous effects are controlled by a length-scale $r_d=\sqrt{\frac{\nu}{\gamma}}$, one expects a crossover from viscous-dominated to relaxation-dominated flow as a function of local channel width $h = r\theta_0$.
In particular, the resistance of the thin $h \ll r_d$ region should scale as $1/h^2$ (Gurzhi/Poiseuille regime), while the resistance of the $h \gg r_d$ region should scale as $1/h$ (Ohmic regime).
Even in the viscous-dominated regime $\gamma \rightarrow 0$, a radial flow assumption is inconsistent with the fixed-voltage boundary conditions as described in the main text; angular components of velocity must contribute.
Therefore, for finite $\nu$ we expect the exact solution of Eq.~\eqref{eq: Venturi Ohmic} to also break down for strong particle flows in the convergent direction, possibly towards turbulence.

\subsection{Purely viscous limit - Jeffrey-Hamel flow}
In the purely viscous limit $\gamma \rightarrow 0$, the leading order flow is a generalization of Poiseuille flow to non-parallel walls. 
This case also admits an exact solution of the Navier-Stokes equation, known as Jeffrey-Hamel flow.\cite{Millsaps1953, RosenheadBook,LandauVol6} 
However, as we are only interested in low-velocity flows, a perturbative treatment will suffice.
In contrast to fixed-voltage boundary conditions, where one cannot assume purely radial flow and therefore is more difficult to solve, we will assume fixed-current boundary conditions where the $\theta$-dependent radial flow $\mathbf{v} = v_r(\theta) \hat{r}$ is a good ansatz.
In addition, we take the divergence-free (incompressible) ansatz $v^{(1)}_r = F(\theta)/r$ for an yet-undetermined function $F$.
On substitution and integration of the $\hat{\mathbf{\theta}}$ NS equation (Eq.~\eqref{eq:Navier-Stokes}), we find that the NS equations give
\begin{align}
    \frac{e}{m} \frac{\partial \phi^{(1)}}{\partial r} =& \frac{\nu}{r^3} \frac{d^2 F}{d\theta^2} \\
    \frac{e}{m}\phi^{(1)} =& \frac{2\nu}{r^2} F(\theta) + S(r)
\end{align}
where $S(r)$ is determined from the boundary conditions.
Substituting for $\phi^{(1)}$, we find that $S(r) = K\frac{\nu}{2r^2} + \text{const}$ for some constant $K$ by separation of variables.
The leading order solution is
\begin{align}
    v^{(1)}_r =& \frac{I}{ne}\frac{1}{r} \frac{1}{\tan\theta_0 - \theta_0}\left(\frac{\cos 2\theta}{\cos \theta_0} - 1\right) \\
    \frac{e}{m}\phi^{(1)} =& \frac{I}{ne} \frac{2\nu}{r^2}\frac{1}{\tan\theta_0 - \theta_0}\frac{\cos 2\theta}{\cos \theta_0}
\end{align}
Since $v^{(2)}_r = 0$, the pressure gradient must balance the convective force.
Therefore, the total potential is given by
\begin{align}
    \frac{e}{m}\phi =& \frac{\nu I}{ne}\frac{1}{r^2}\frac{1}{\tan\theta_0 - \theta_0} \left(\frac{\cos 2\theta}{\cos\theta_0} + \frac{I}{2n e\nu}\frac{1}{\tan\theta_0 - \theta_0}\left(\frac{\cos 2\theta}{\cos\theta_0}-1\right)^2\right)
\end{align}
We see that $\phi^{(2)}$ is suppressed by a viscous Reynolds number $\operatorname{Re}_\nu \sim \frac{I}{ne\nu}$, as expected.
Analogous to the purely Ohmic case discussed in the main text, it is known that divergent Jeffrey-Hamel flow is unstable towards turbulence.\cite{LandauVol6,RosenheadBook}

\section{Eckart Streaming}
\label{sec: appendix Eckart}
In this section, we lay out the mathematical calculation of Sec.~\ref{sec: Eckart} in full detail.

\subsection{Leading order solution}

As mentioned in the main text, we take the ansatz that the leading order solution is described by a longintudinal sound mode with wavevector $k_l = k + i\alpha$ (see Eq.~\eqref{eq:longitudinal dispersion}).
Applying the voltage-fixed boundary conditions and using the density continuity equation (see Eq.~\eqref{eq:density continuity}), we find
\begin{align}
    \phi^{(1)} =& V_0\Re\left[\frac{e^{(ik -\alpha)x} - e^{(ik-\alpha)(2l-x)}}{1-e^{(ik-\alpha) 2l}} e^{-i\omega t}\right] \\
    v_x^{(1)} =& u_0 \Re\left[\frac{e^{(ik -\alpha)x} + e^{(ik-\alpha)(2l-x)}}{1-e^{(ik-\alpha) 2l}} e^{-i\operatorname{Arg} k_l} e^{-i\omega t}\right]
\end{align}
where $u_0 = \frac{CV_0}{\rho_e^{(0)}} \frac{\omega}{|k_l|}$ and $\Re$ denotes real part.
To satisfy the no-slip boundary, we must also include a divergence-free (incompressible) contribution to the flow corresponding to a boundary layer correction, as is done in Sec.~\ref{sec: Rayleigh}. 
We defer the discussion of this correction to the end of this section, assuming that its contribution is small.

\subsection{Second-order density continuity equation}
We now turn to the time-averaged second-order hydrodynamic equations, where we have assumed $\overline{v^{(2)}_y} = 0$.
The density continuity (i.e. current conservation) equation (see Eq.~\eqref{eq:density continuity}) gives
\begin{align}
    \frac{\partial \overline{J_x^{(2)}}}{\partial x} \equiv \frac{\partial}{\partial x} \left[\rho_e^{(0)}\overline{v_x^{(2)}} + \overline{\rho_e^{(1)} v_x^{(1)}}\right] = 0
\end{align}
which tells us that $\overline{J_x^{(2)}}(y)$ only depends on $y$.
We remark that it is crucial that $\overline{\mathbf{v}^{(2)}}$ is \emph{not} divergence-free (incompressible); because the ``drift'' contribution $\overline{\rho_e^{(1)} v_x^{(1)}}$ is non-zero and $x$-dependent, divergence-ful (compressive) contributions of $v_x^{(2)}$ are necessary to satisfy current conservation.

\subsection{Second-order Navier-Stokes equation - DC forces and screening}
\label{sec: appendix Eckart forces}
Replacing $\overline{v_x^{(2)}}$ in favor of $\overline{J_x^{(2)}}$ in the Navier-Stokes equation (see Eq.~\eqref{eq:Navier-Stokes}), we get
\begin{align}
    \frac{m}{e}\left[-\nu\frac{\partial^2}{\partial y^2} + \gamma\right] \overline{J_{x}^{(2)}} =& \overline{F_\text{eff}^{(2)}}
    \label{eq:appendix Eckart Poiseuille}
    \\
    -\rho_e^{(0)} \frac{\partial \overline{\phi_\text{eff}^{(2)}}}{\partial x} + \overline{F^{(2)}_{\text{conv},x}} + \overline{F^{(2)}_{\text{elec},x}} + \overline{F^{(2)}_{\text{comp},x}} \equiv&  \overline{ F_\text{eff}^{(2)}}
    \label{eq:appendix Eckart forces}
\end{align}
where we used separation of variables with constant $\overline{F_\text{eff}}$ to split the momentum equation, and $\rho_e^{(0)}\phi_\text{eff}^{(2)} \equiv \rho_e^{(0)} \phi^{(2)} + p^{(2)}$.
We remark that Eq.~\eqref{eq:appendix Eckart Poiseuille} is an Ohmic-Poiseuille equation\cite{Torre2015a} describing steady, divergence-free (incompressible) flow in rectangular channel, where $\overline{F_\text{eff}}$ can be interpreted as the effective force driving the flow.
The convective force is defined in Eq.~\eqref{eq:F_conv}, while the terms $\overline{F_{\text{elec},x}^{(2)}}$ and $\overline{F_{\text{comp},x}^{(2)}}$ are given by
\begin{align}
    \overline{F_{\text{elec},x}^{(2)}} =& \overline{\rho_e^{(1)}\frac{\partial \phi^{(1)}}{\partial x}}
    \\
    \overline{F_{\text{comp},x}^{(2)}} =& (2\nu + \tilde{\zeta})\left[\overline{\rho^{(1)}\frac{\partial^2 v_x^{(1)}}{\partial x^2}} - \overline{\frac{\partial^2 \left(\rho^{(1)}v_x^{(1)}\right)}{\partial x^2}}\right] 
\end{align}
where in the second line we have used $\frac{\partial}{\partial x} (\overline{\rho^{(0)} v_x^{(2)}}) = - \frac{\partial}{\partial x} (\overline{\rho^{(1)} v_x^{(1)}})$.
These provide nonlinear contributions to $\overline{F_\text{eff}^{(2)}}$ in addition to the convective force.
The first term comes from the backreaction of the electric force; we remark that the presence of this nonlinearity was also noted by Ref.~\onlinecite{Torre2015}.
The second term comes from compressive dissipation. 
By solving for $\overline{\phi_\text{eff}^{(2)}}$ with the zero-voltage boundary conditions, we find the simple result
\begin{align}
    \overline{F_\text{eff}} = \frac{1}{l}\int_0^l dx \overline{F^{(2)}_{\text{conv},x}} + \overline{F^{(2)}_{\text{elec},x}} + \overline{F^{(2)}_{\text{comp},x}}
    \label{eq:appendix Eckart screening}
\end{align}
The action of the effective voltage is to ``screen'' all the forces via a spatial average, rendering the resulting effective force $x$-independent. 
We comment that $\frac{1}{l}\int_0^l dx\overline{F_{\text{elec},x}^{(2)}} = \frac{CV_0^2}{4l}$ has no $\alpha$ or $k$ dependence, and therefore no interference behavior; the value of $\overline{ F_{\text{elec},x}^{(2)}}$ is fixed at the ends by the voltage boundary conditions. 
By dimensional analysis, these contributions are small relative to the convective force when $\frac{s_\text{cap}^2\omega^2}{|k_l|^2} \ll 1$ and $\frac{(2\nu + \tilde{\zeta})|k_l|^2}{\omega} \ll 1$, respectively.
For parameters as discussed in the main text, we find $\frac{s_\text{cap}^2|k_l|^2}{\omega^2} \sim .24$ and $\frac{(2\nu + \tilde{\zeta}) |k_l|^2}{\omega} \sim .06$ are small, so that ignoring $\overline{F_{\text{elec},x}^{(2)}}$ and $\overline{F_{\text{comp},x}^{(2)}}$ is valid.

\subsection{Rectified DC solution}
The solution of the Ohmic-Poiseuille equation (Eq.~\ref{eq:appendix Eckart Poiseuille}) is
\begin{align}
    \overline{J_{x}^{(2)}} =& \rho_e^{(0)} u_0 \left[\frac{\overline{F_\text{eff}^{(2)}}}{\rho^{(0)}\gamma u_0}\right] \left(1 -  \frac{\cosh\frac{y}{r_d}}{\cosh{\frac{h}{2r_d}}}\right) \\
    \overline{I^{(2)}}  =& I_0 \left[\frac{\overline{F_\text{eff}^{(2)}}}{\rho^{(0)}\gamma u_0}\right] \left(1 - \frac{2r_d}{h} \tanh \frac{h}{2r_d} \right)
\end{align}
The term in square brackets is suggestively written to resemble momentum-relaxation Reynolds number $\operatorname{Re}_\gamma$, which is indeed true when the convective force dominates (see Eq.~\eqref{eq:Eckart Reynolds number}).
We remark that the convective contribution to $I^{(2)}/I_0$ is largely $\alpha$-independent (see Eq.~\eqref{eq:Eckart current}); in the limit $\alpha l \gg 1$, where the interference term can be neglected, the result is surprisingly $\alpha$-independent even though $\alpha$ was necessary to generate convective gradients.
Instead, the scale of the convective gradient is screened, being controlled by the device length $l^{-1}$.
This $\alpha$-independence has an analogue in Rayleigh streaming, where the shear viscosity $\nu$ does not set the scale of the rectified bulk flow even though it was necessary to set up convective forces.

\subsection{Revisiting Boundary Dissipation (Rayleigh Streaming)}
\label{sec: appendix Eckart boundary dissipation}

We return to the issue of the no-slip condition and boundary layer corrections (i.e Rayleigh streaming), which we ignored for the leading order solution.
For simplicity, we will neglect contributions from the reflected wave (i.e. $\alpha l \gg 1$).
As discussed in Sec.~\ref{sec: Rayleigh}, boundary layer corrections are described by the transverse mode $k_t = k_t' + ik_t''$, decaying exponentially from the wall with length $1/k_t''$.
For parameters as discussed in the main text, we find $k_t'' h \sim 8.2 > 1$ so that it is a good assumption that the boundary layer is thin.
Therefore, boundary dissipation (i.e. Rayleigh streaming) effects will lead to a non-zero slip velocity for the bulk flow also in the forward $x$-direction. 
Upon solving the Ohmic-Poiseuille equation (Eq.~\eqref{eq:appendix Eckart Poiseuille}) with a voltage-fixed boundary condition $\phi(x=l)=0$ (as in the main text), we get an additional contribution
\begin{align}
    \overline{J_{\text{Rayleigh}, x}^{(2)}} =& v^{(2)}_\text{slip}\frac{\cosh \frac{y}{r_d}}{\cosh \frac{h}{2r_d}}
    \\
    \overline{I_{\text{Rayleigh}}^{(2)}} =& v^{(2)}_\text{slip}\tanh \frac{h}{2r_d}
\end{align}
Therefore, the no-slip boundary (i.e. Rayleigh streaming) only provides a quantitative correction to the DC current.
By estimating $v_\text{slip}^{(2)} \sim u_0 e^{-1/k_t'' r_d} \frac{I_0 |k_l|}{\rho_e^{(0)} h \gamma}$ from the Rayleigh Reynolds number in Eq.~\eqref{eq:Rayleigh Reynolds number} with exponential decay arising from the viscous length scale $r_d$, we find that boundary dissipation contributes additively to the bulk dissipation contribution.

If instead one takes the current-fixed boundary condition $J(x=l) = 0$, a rectified DC voltage will develop as discussed in previous works.\cite{Dyakonov1996, Torre2015}
However, these previous works did not consider the effect of a no-slip boundary.
As a result of no-slip, we expect only a quantitative change to the DC voltage analogous to the previous case.
However, a qualitative change occurs in the current flow - a circulating current must develop in the channel as in Sec.~\ref{sec: Rayleigh}.
The length and width of the circulation will be set by the device dimensions, as opposed that of Sec.~\ref{sec: Rayleigh} where the length is set by the wavelength.
Surprisingly, the bulk current density flows in an opposite direction to that of the convective force; because convective forces are stronger near the boundary than the bulk, the forward DC flow along $x$ must be near the boundary while the counterflow is in the bulk.\cite{Nyborg1965}
This reversed bulk counterflow would be also be interesting evidence for hydrodynamic behavior, though measuring the local current density may prove challenging.

\section{Rayleigh Streaming}
\label{sec: appendix Rayleigh}
In this section, we fill out the mathematical details of Sec.~\ref{sec: Rayleigh}.

\subsection{Leading order solution - Boundary corrections}
Recall that we work in the limit $k_t''^{-1} \ll h \ll \lambda$ of a thin boundary layer and long wavelength.
In this limit, we can separate the flow into bulk and boundary regions, stitching the flow together at the interface.
We first focus on the boundary layer region, concentrating on the lower boundary layer near $y=-h/2$; flow at the upper boundary layer is given by reflection symmetry about $y=0$.
In the lower boundary layer, the leading-order longitudinal (irrotational) and transverse (incompressible) velocity components of $\mathbf{v}_\text{wall}^{(1)}$ are
\begin{align}
    v_{\text{wall},l,x}^{(1)} =& v_{l,x}^{(1)} = u_0 \sin kx \Re e^{i\omega t}
    \\
    v_{\text{wall},t,x}^{(1)} =& -u_0 \sin kx \Re\left[e^{ik_t y_+}e^{-i\omega t}\right]
    \\
    v_{\text{wall},t,y}^{(1)} =& -u_0 k \cos kx \Re\left[\left(1 - e^{ik_t y_+} \right)\frac{e^{-i\omega t}}{ik_t}\right]
\end{align}
where $y_+ = y+\frac{h}{2}$ is the distance from the lower wall, we take $k_t''>0$, and $\Re$ denotes real part.
Although $v_{\text{wall},y}^{(1)}$ is small compared to $v_{\text{wall},x}^{(1)}$, the $y$-gradients of $v_{\text{wall},y}^{(1)}$ are large and must be included when computing the convective force.
The longitudinal contribution $v_{\text{wall},l,x}^{(1)}$ is inherited from the longitudinal ansatz of Eq.~\eqref{eq:Rayleigh ansatz}.
We remark that we have not assumed that $v_\text{wall}^{(1)}$ is divergence-free (incompressible) unlike classic discussions\cite{Rayleigh1883,LandauVol6,Riley2001}; that the divergence-free (incompressible) ansatz is not correct has been previously pointed out,\cite{Westervelt1953, Nyborg1965}, though it has no consequence in the limit $\gamma \rightarrow 0$.
In the limit $k_t'' y_+ \gg 1$, we find that $v_{\text{wall},x}$ returns to our longitudinal ansatz $v_{l,x}^{(1)}$ as the boundary-layer corrections exponentially vanish.
However, $v_{\text{wall},t,y}^{(1)}$ is non-zero in this limit and requires correction in the bulk.
We will not concern ourselves with the bulk corrections to $v^{(1)}_y$, as they are small and do not contribute substantially to the convective force.\cite{Nyborg1965}

Therefore, the convective force in the bulk and boundary layers are
\begin{align}
    \overline{F_{\text{conv, bulk},x}^{(2)}} =& \rho^{(0)} u_0^2 k\frac{1}{4} \sin 2kx (-2)
    \\
    \overline{F_{\text{conv, wall},x}^{(2)}} =& \rho^{(0)} u_0^2 k\frac{1}{4} \sin 2kx \left[-2 + (3 + e^{2i\theta_t}) e^{ik_t y_+} - 2e^{-2k_t'' y_+}\cos^2{\theta_t}\right]
\end{align}
where $\theta_t \equiv \operatorname{Arg} k_t$.

\subsection{Second-order Navier-Stokes}
We now study the DC second-order flow.
We begin by noting that the assumption $k_t''^{-1} \ll h \ll \lambda$ implies that $v_y \ll v_x$, i.e. flow is primarily along $x$ because the channel is thin.
By using the NS equations (Eq.~\eqref{eq:Navier-Stokes}), this implies that the effective voltage $\phi_\text{eff} = \phi + \frac{1}{\rho_e^{(0)}} p$ satisfies $\frac{\partial \phi_\text{eff}}{\partial y} \ll \frac{\partial \phi_\text{eff}}{\partial x}$, i.e. voltage gradients (and density gradients) are also primarily along $x$.

Next, we simplify the NS equation (Eq.~\eqref{eq:Navier-Stokes}).
First, we note that the backreactive electric force $\mathbf{F}_\text{elec}^{(2)} \equiv \overline{\rho_e^{(1)}\nabla \phi^{(1)}} = 0$.
We will also assume that compressional dissipation $\mathbf{F}_\text{comp} \equiv (2\nu+\tilde{\zeta})\rho\nabla\nabla\cdot \mathbf{v}$ is negligible, which is consistent with our assumption that the longitudinal attenuation $\alpha$ is small.
Finally, for simplicity we neglect the additional term $\nu \overline{\rho_e^{(1)} \nabla\times\nabla\times \mathbf{v}^{(1)}}$ as is done in classical treatments of Rayleigh streaming;\cite{Rayleigh1883, LandauVol6, Nyborg1965, Westervelt1953, Riley2001} this term depends on the density dependence of $\nu$, where classical works assumed that the dynamic viscosity $\mu\equiv \rho \nu$ is constant.
Therefore, the NS equation becomes
\begin{align}
    \frac{m}{e}\left[-\nu\frac{\partial^2}{\partial y^2} + \gamma\right] \overline{J_x^{(2)}} =&  \overline{F^{(2)}_{\text{conv},x}} - \rho_e^{(0)} \frac{\partial \overline{\phi_\text{eff}^{(2)}}}{\partial x}
    \label{eq:appendix Rayleigh Poiseuille}
\end{align}
where we have used $k_t''^{-1} \ll h \ll \lambda$ to drop the $x$-derivatives (cf. Eq.~\eqref{eq:appendix Eckart Poiseuille} and Eq.~\eqref{eq:appendix Eckart forces}).
Note that this form is equivalent to assuming that $\mathbf{v}^{(2)}$ is divergence-free (incompressible).

Since the convective force is only $x$-dependent in the bulk, we must have
\begin{align}
    \rho_e^{(0)} \frac{\partial \overline{\phi_\text{eff}^{(2)}}}{\partial x} =& \overline{F_{\text{conv,bulk},x}}
\end{align}
upon imposing $\overline{I}=0$ (i.e. $J_x(x=0) = 0$).
More concretely, the boundary conditions for $\overline{v_x^{(2)}}(y=\pm h/2)$ will fix the $y$-dependent homogeneous solutions of Eq.~\eqref{eq:appendix Rayleigh Poiseuille}, leaving $\overline{\phi_\text{eff}^{(2)}}$ to enforce $\overline{I^{(2)}}=0$. 
Since $\frac{\partial \overline{\phi_\text{eff}^{(2)}}}{\partial y}$ is small, this expression for $\phi_\text{eff}^{(2)}$ is also valid in the boundary layer.
Therefore, after ``screening'' from the effective voltage, the resultant force is only non-zero in the boundary layer.

\subsection{Second-order boundary layer solution}
We first solve Eq.~\eqref{eq:appendix Rayleigh Poiseuille} in the boundary layer, where the ``screened'' convective force is not negligible.
Assuming $r_d\ll h$, the solution for the lower boundary layer is
\begin{align}
    \overline{J_{\text{wall},x}^{(2)}} =& \rho_e^{(0)}u_0 \sin 2kx \, \Re\left[\frac{v_\text{slip}}{u_0}  - \frac{u_0 k}{4\gamma} \left(-\frac{(3+e^{2i\theta_t})e^{ik_t y_+}}{k_t^2 r_d^2 + 1} - \frac{(2 \cos^2\theta_t) e^{-2k_t'' y_+}}{4 k_t''^2 r_d^2 - 1}\right)\right]
    \\
    v_\text{slip}^{(2)}=& \frac{u_0^2 k}{4\gamma} e^{-\frac{y_+}{r_d}} \Re\Bigg[-\frac{(3+e^{2i\theta_t})(i\tilde{\omega}+2)}{4+\tilde{\omega}^2} 
    - \frac{2\cos^2\theta_t}{-3+2\sqrt{1+\tilde{\omega}^2}}\Bigg] 
\end{align}
where $v_\text{slip}^{(2)}$ enforces the no-slip boundary conditions and have rewritten $k_t^2 r_d^2$ in terms of $\tilde{\omega}=\omega/\gamma$ using Eq.~\ref{eq:transverse dispersion}.
Away from the wall where the convective force vanishes, the velocity $v_{\text{wall},x}^{(2)} \rightarrow v_\text{slip}^{(2)} \sin 2kx$ achieves a non-zero limiting value if $k_t'' r_d$ is sufficiently large; the boundary layer sets up a slip boundary for the bulk flow.
In the main text, we (optimistically) approximate the size of the boundary to be $1/k_t''$ so that we evaluate $v_\text{slip}^{(2)}$ at $y_+/r_d = 1/(k_t'' r_d)$.
The resulting bulk flow is solved from Eq.~\eqref{eq:appendix Rayleigh Poiseuille} with a vanishing RHS and with the slip boundary generated from the boundary layer; the solution is given in the main text (Eq.~\eqref{eq:Rayleigh vx} and Eq.~\eqref{eq:Rayleigh vy}).

We make three remarks on $v_\text{slip}$.
First, in the limit $\nu \rightarrow 0$, the flow becomes increasingly singular at the walls so the boundary layer will no longer by described by hydrodynamics.
Second is the surprising fact that $\nu$ is largely $\nu$-independent.
In the limit $\gamma\rightarrow 0$, we recover the classical result $v_\text{slip} = -\frac{3 u_0}{8}\frac{u_0 k}{\omega}$ which is $\nu$-independent, despite the necessity of $\nu$ to set up convective gradients.
Instead of the viscous Reynolds number $\operatorname{Re}_\nu$, the slip velocity is controlled by the Mach number $u_0\omega/k$.
This was first noted by Rayleigh in the classical situation.\cite{Rayleigh1883}
Finally, $v_\text{slip}$ has a resonance at $\omega = \frac{\sqrt{5}}{2}\gamma$ corresponding to $-4k_t''^2 r_d^2 +1 = 0$.
We leave further study of this interesting convective instability to future work; for this paper we only work in the limit $v_\text{slip}\ll u_0$ where perturbation theory is valid.

\end{document}